\documentclass[review]{elsarticle}

\usepackage{lineno,hyperref}
\usepackage{natbib}
\modulolinenumbers[5]

\journal{Journal of \LaTeX\ Templates}

\begin{document}

\begin{frontmatter}

\title{Constraining the optical potential in the search for $\eta$-mesic 
$^{4}\hspace{-0.03cm}\mbox{He}$}  




\author[uj]{M. Skurzok}
\ead{magdalena.skurzok@uj.edu.pl}

\author[uj]{P. Moskal}

\author[andes]{N. G. Kelkar}
\author[nara]{S. Hirenzaki}
\author[nara,osaka]{H. Nagahiro}
\author[totto]{N. Ikeno}

\address[uj]{Institute of Physics, Jagiellonian University, prof. Stanis{\l}awa {\L}ojasiewicza~11, 30-348 Krak\'{o}w, Poland}
\address[andes]{Departamento de Fisica, Universidad de los Andes, Cra.~1E, 18A--10, Bogot{\'a}, Colombia}
\address[nara]{Department of Physics, Nara Women's University, Nara 630-8506, Japan}
\address[osaka]{Research Center for Nuclear Physics (RCNP), Osaka University, Ibaraki 567-0047, Japan}
\address[totto]{Department of Life and Environmental Agricultural Sciences, 
Tottori University, Tottori 680-8551, Japan} 
\begin{abstract}
A consistent description of the $ d d \rightarrow$ $^{4}\hspace{-0.03cm}\mbox{He}  
\eta$ and $ d d \rightarrow$ ($^4$He$\eta$)$_{bound} \rightarrow X $  cross sections 
was recently proposed with a broad range of real ($V_{0}$) and imaginary ($W_{0}$), 
$\eta$-$^4$He optical potential parameters leading to a good 
agreement with the $ d d \rightarrow$ $^{4}\hspace{-0.03cm}\mbox{He} \eta$ data. 
Here we compare the predictions of the model below the $\eta$ production threshold, with 
the WASA-at-COSY excitation 
functions for the $dd\rightarrow$ $^{3}\hspace{-0.03cm}\mbox{He} N \pi$ reactions to 
put stronger constraints on $(V_0, W_0)$. 
The allowed parameter space (with $|V_0| <\, \sim$ 60 MeV and $|W_0| <\, \sim$ 
7 MeV estimated at 90\% CL ) excludes
most optical model predictions of $\eta-^4$He nuclei except for  
some loosely bound narrow states.
\end{abstract}


\end{frontmatter}


\section{Introduction}

\label{intro}

\textit{Mesic nuclei} are currently one of the hottest topics in nuclear and hadronic physics, both from experimental~\cite{Skurzok_NPA,Adlarson_2013,Tanaka,Machner_2015,Metag2017} and theoretical points of view~\cite{Ikeno_EPJ2017,Xie2017,Fix2017,Barnea2017,Gal2017,Gal2015,Friedman,Kelkar_2016_new,Kelkar,Kelkar_new,Wilkin_2016,Wilkin2,BassTom2006,BassTom,Hirenzaki1,Nagahiro_2008rj,Nagahiro_2013,Hirenzaki_2010,WycechKrzemien,Niskanen_2015}. This exotic nuclear matter is supposed to consist of a nucleus bound via the strong interaction with a neutral meson such as the $\eta$, $\eta'$, $K$ or $\omega$. Although, its existence has been predicted over thirty years ago, it still remains to be one of the undiscovered nuclear objects. Some of the most promising candidates for such bound states are $\eta$-mesic nuclei, postulated by Haider and Liu in 1986~\cite{HaiderLiu1} following the coupled channel calculations by Bhalerao and Liu~\cite{BhaleraoLiu} which reported an attractive $\eta$-nucleon interaction. Current studies of hadron- and photon-induced production of the $\eta$ meson resulting in a wide range of values of the $\eta N$ scattering length, $a_{\eta N}$, indicate the interaction between the $\eta$ meson and a nucleon to be attractive and strong enough to create an $\eta$-nucleus bound system even in light nuclei~\cite{Xie2017,Fix2017,Barnea2017,Gal2017,Green,Wilkin1,WycechGreen}. However, experiments performed so far have not brought a clear evidence of their 
existence~
\cite{Berger,Mayer,Sokol_2001,Smyrski1,Mersmann,Budzanowski,Papenbrock,PMActa}. 
They provide only signals which might be interpreted as indications of the 
$\eta$-mesic nuclei. 
The interested reader can find recent reviews on the $\eta$ mesic bound states searches in Refs~\cite{Machner_2015,Metag2017,Kelkar,Wilkin_2016,HaiderLiu,Krusche_Wilkin,Moskal_FewBody,Acta_2016,Moskal_Acta2016,Moskal_AIP2017}. 

Some of the promising experiments related to $\eta$-mesic nuclei have been performed 
with the COSY facility~\cite{Wilkin_epj2017}. 
The most recent of these involves the measurement of the 
$dd\rightarrow$ $^{3}\hspace{-0.03cm}\mbox{He} n \pi^{0}$ 
and $dd\rightarrow$ $^{3}\hspace{-0.03cm}\mbox{He} p \pi^{-}$ 
reactions which has been performed by the WASA-at-COSY Collaboration. Due to the lack of theoretical predictions for cross sections below the $\eta$ production threshold, the data have been analyzed assuming that the signal from the bound state has a Breit-Wigner shape~\cite{Skurzok_NPA,Adlarson_2013}. However, a better guidance for the shape of the cross sections for the $dd\rightarrow$ ($^{4}\hspace{-0.03cm}\mbox{He}$-$\eta)_{bound} \rightarrow$ $^{3}\hspace{-0.03cm}\mbox{He} N \pi$ processes is provided by a theoretical model described in 
Ref.~\cite{Ikeno_EPJ2017} in the excess energy range relevant 
to the $\eta$-mesic nuclear search. Given that the model is the very first attempt to provide a consistent description of the data below and above the $\eta$ meson production threshold, the authors used a phenomenological approach with an optical potential for the $\eta$-$^4$He interaction. The available data on the $d d \rightarrow \, ^4$He $\eta$ reaction is reproduced quite well for a broad range of optical potential parameters for which the authors predict the cross section spectra corresponding to $\eta$-$^4$He bound state formation in the subthreshold region. 
In this article we present a comparison between this new theoretical model and experimental data collected by WASA-at-COSY in order to further constrain the range of the 
allowed $\eta$-$^4$He optical potential parameters. The latter, as we shall see, narrows 
down the search for $\eta$-mesic helium to a region of small binding energies 
and widths.


\section{Theoretical model}

\begin{figure}[h!]
\centering
\includegraphics[width=8.0cm,height=5.3cm]{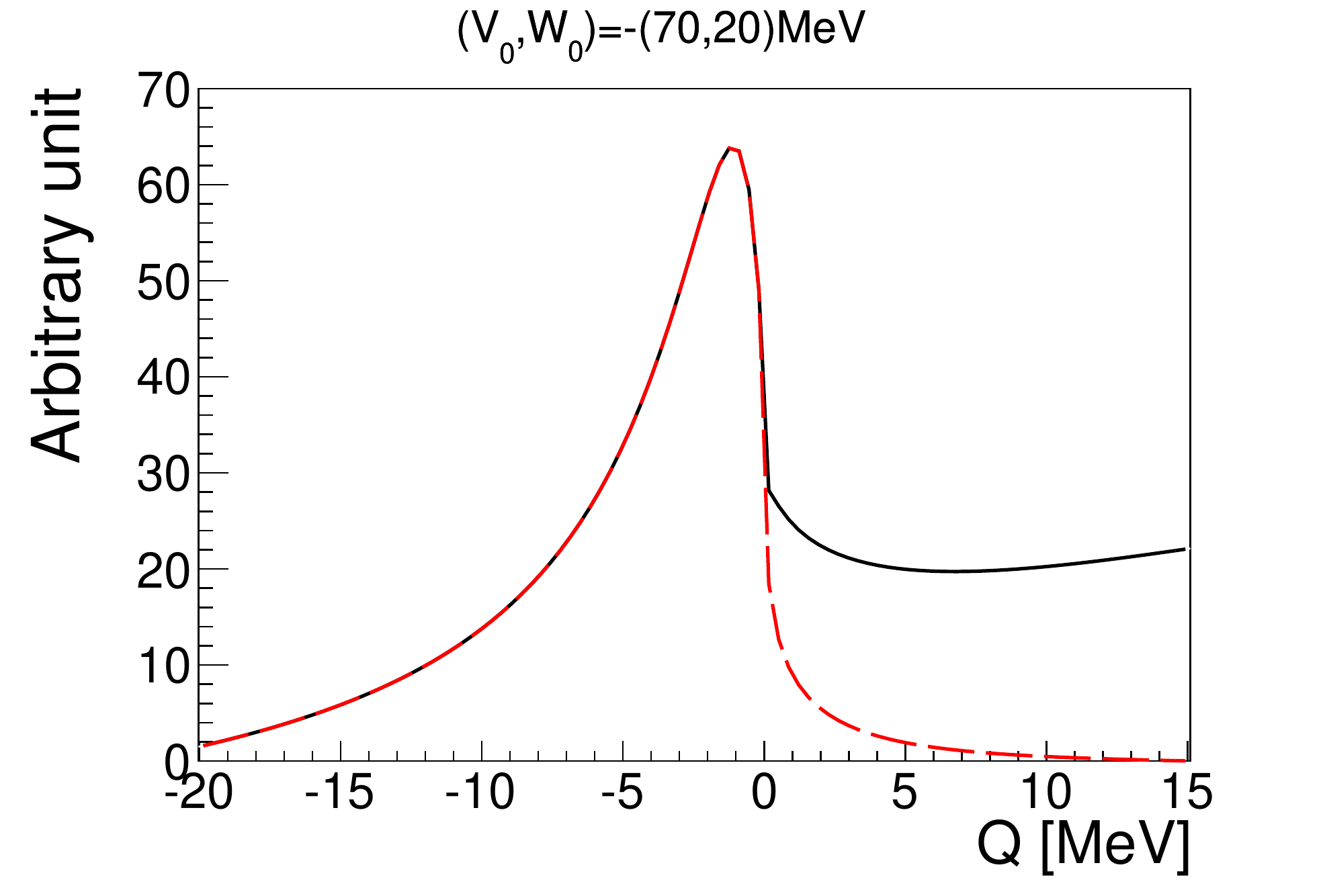}
\caption{Calculated total cross section of the $dd\rightarrow$ ($^{4}\hspace{-0.03cm}\mbox{He}$-$\eta)_{bound}  \rightarrow$ $^{3}\hspace{-0.03cm}\mbox{He} N \pi$ reaction for the formation of the $^{4}\hspace{-0.03cm}\mbox{He}$-$\eta$ bound system plotted as function of the excess energy Q for $\eta$-$^{4}\hspace{-0.03cm}\mbox{He}$ optical potential parameters ($V_{0},W_{0}$)=$-$(70,20)~MeV. The black solid line denotes the 
total cross section $\sigma$, while the red dashed line denotes the conversion part $\sigma_{conv}$.~\label{theory_total}}
\end{figure}

The  formalism presented in Ref.~\cite{Ikeno_EPJ2017} predicted for the first time, 
the formation rate of the $\eta$-mesic $^{4}\hspace{-0.03cm}\mbox{He}$ in the 
deuteron-deuteron fusion reaction within a model which reproduced the data on 
the $ d d \rightarrow \, ^4$He $\eta$ reaction quite well. The authors determined the total cross sections for the $dd\rightarrow$ ($^{4}\hspace{-0.03cm}\mbox{He}$-$\eta)_{bound} \rightarrow$ $^{3}\hspace{-0.03cm}\mbox{He} N \pi$ reaction based on phenomenological calculations. The calculated total cross section $\sigma$ consists of two parts: conversion $\sigma_{conv}$ and escape $\sigma_{esc}$ part. The conversion part, determined for different parameters $V_{0}$ and $W_{0}$ of a spherical $\eta$-$^{4}\hspace{-0.03cm}\mbox{He}$ optical potential $V(r) = (V_{0} + iW_{0}) \frac{\rho_{\alpha}(r)}{\rho_{\alpha}(0)}$, is equal to the total cross section in the subthreshold excess energy region where the $\eta$ meson is absorbed by the nucleus (its energy is not enough to escape from the nucleus), while the $\eta$ meson escape part contributes to the excess energy region above the threshold for $\eta$ production and can be calculated as $\sigma_{esc}=\sigma-\sigma_{conv}$. Fig.~\ref{theory_total} shows the example of a calculated total cross section for $\eta$-$^{4}\hspace{-0.03cm}\mbox{He}$ optical potential parameters ($V_{0},W_{0}$)=$-$(70,20)~MeV.

We should mention here that the above theoretical calculations 
(which are being used in the present work) were 
done assuming the one-nucleon absorption of the $\eta$ meson since 
the strength of the multi-nucleon absorption processes is not well known. 
Based on the experimental data on the $ p n \to d \eta$ and $ p N \to p N \eta$ 
reactions, the strength of the $\eta$ meson absorption by a two-nucleon pair at the 
nuclear center was estimated in \cite{WycechAPPB29} 
to be 4.2 MeV and 0.2 MeV for the spin triplet and singlet nucleon pairs, 
respectively. 
This strength can be larger for $^4$He because of the higher central density 
as mentioned in \cite{WycechAPPB45}.  
The values of the $W_{0}$ parameters in the present work 
could be compared with these numbers  
to get a rough estimate of the ratio of the one- and two-body absorption 
probability at the nuclear center. 
The two body absorption potential is expected to provide an additional contribution 
to the conversion cross section. 
However, it is only the one-nucleon absorption cross section which should be 
compared with the present data since multi-nucleon absorption processes 
would contribute to different final states not considered in the present work.
Thus, the present analysis of experimental data from 
Ref. \cite{Skurzok_NPA} based on the theoretical calculation assuming 
the one-body absorption seems reasonable.


\begin{figure}[h!]
\centering
\includegraphics[width=8.0cm,height=5.3cm]{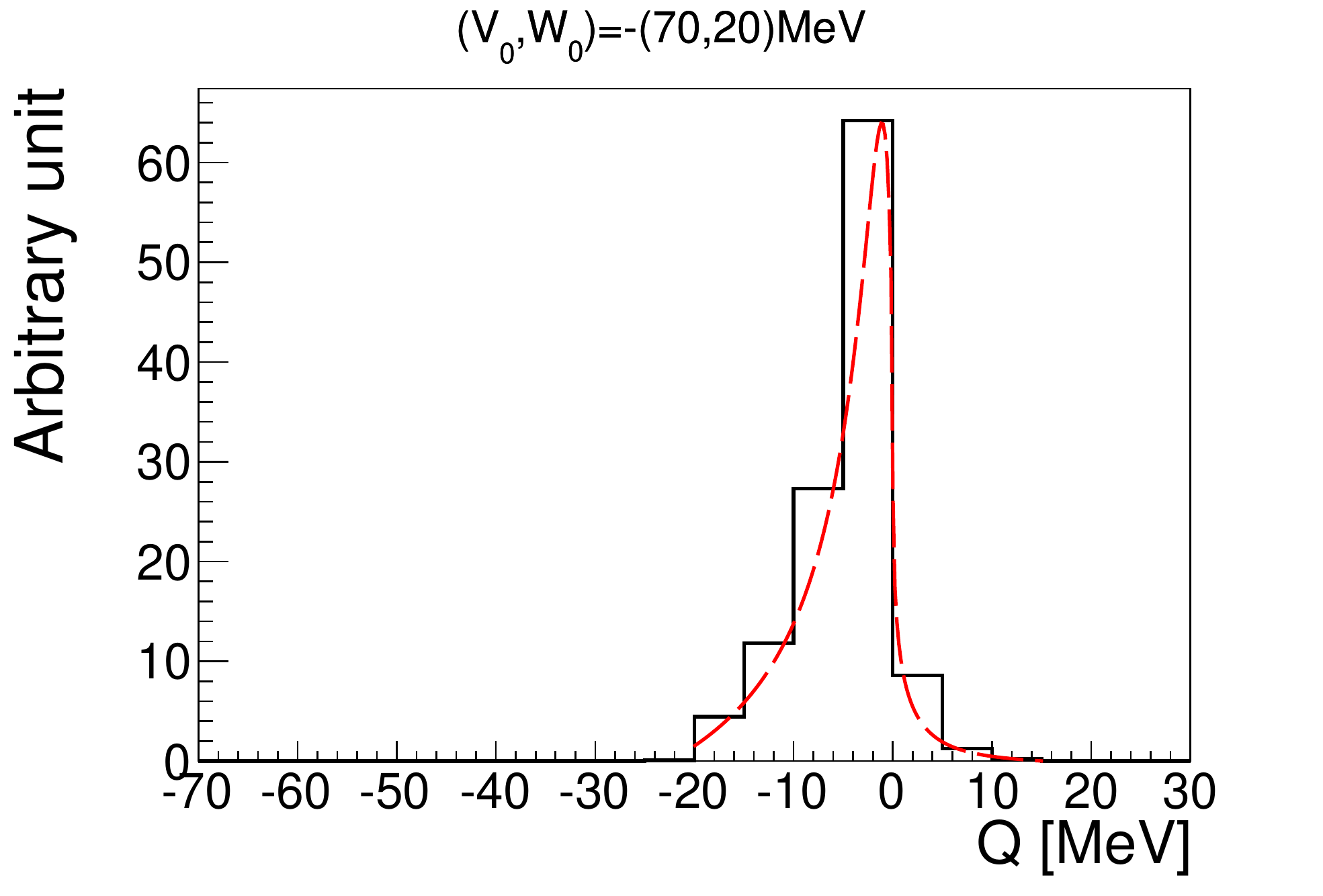}
\caption{Calculated conversion part of the cross section of the $dd\rightarrow$ ($^{4}\hspace{-0.03cm}\mbox{He}$-$\eta)_{bound} \rightarrow$ $^{3}\hspace{-0.03cm}\mbox{He} N \pi$ reaction for the formation of the $^{4}\hspace{-0.03cm}\mbox{He}$-$\eta$ bound system plotted as a function of the excess energy Q for $\eta$-$^{4}\hspace{-0.03cm}\mbox{He}$ optical potential parameters ($V_{0},W_{0}$)=$-$(70,20)~MeV. The cross section is scaled by fitting the escape part to the existing $dd\rightarrow$ $^{4}\hspace{-0.03cm}\mbox{He}\eta$ data and the flat contribution is subtracted as well. The red dashed line shows the theoretical spectrum while the black solid line shows the spectrum after binning (details in Sec.~\ref{Sec_2}).~\label{theory_total_norm}}
\end{figure}

The spectrum has been normalized in the sense that the escape part reproduces the measured cross sections for the $dd\rightarrow$ $^{4}\hspace{-0.03cm}\mbox{He}\eta$ process~\cite{Frascaria,Willis,Wronska}. Moreover, the flat contribution in the conversion spectrum, considered to be a part of the background, has been subtracted (taking minimum value of the $\sigma_{conv}$ in the excess energy range from -20 to 15~MeV). \\
\indent Since the signal from the $\eta$-mesic bound system is expected below the threshold for the $\eta$ meson production, authors focused here only on the conversion part of the cross sections. An example of the calculated $\sigma_{conv}$ is shown in Fig.~\ref{theory_total_norm} for potential parameters ($V_{0},W_{0}$)=$-$(70,20)~MeV. 
Authors of Reference~\cite{Ikeno_EPJ2017} concluded that as a next step it would be important to compare these theoretical results with the experimental data, convoluting the theoretical cross sections with the experimental resolution functions. In this article we present results of such a comparison. The details are presented in Section~\ref{Sec_3} which will be preceded by a brief description of the experimental conditions. 



\section{Experimental data}
\label{Sec_2}

Recent measurements at WASA-at-COSY, dedicated to search for $\eta$-mesic $^{4}\hspace{-0.03cm}\mbox{He}$ nuclei were carried out using the unique ramped beam technique allowing for the beam momentum to be changed slowly and continuously around the $\eta$ production threshold in each of the acceleration cycles~\cite{Skurzok_NPA,Adlarson_2013,Acta_2016,Moskal_AIP2017}. This technique allows to reduce systematic uncertainties with respect to separate runs at fixed beam energies~\cite{Adlarson_2013,Smyrski1,Moskal_1998}.
The $^{4}\hspace{-0.03cm}\mbox{He}$-$\eta$ bound states were searched by studying the excitation functions for \mbox{$dd\rightarrow$ $^{3}\hspace{-0.03cm}\mbox{He} n \pi^{0}$} and \mbox{$dd\rightarrow$ $^{3}\hspace{-0.03cm}\mbox{He} p \pi^{-}$} processes in the excess energy range $Q$ from -70~MeV to 30~MeV. 
The obtained excitation functions do not reveal any direct narrow structure below the $\eta$ production threshold, which could be considered as a signature of the bound state. Therefore, only the upper limit of the total cross section for the $\eta$-mesic $^{4}\hspace{-0.03cm}\mbox{He}$ formation was determined. 

In the first approach, the upper limits of the total cross sections for both processes were estimated at a 90\% confidence level (CL) fitting simultaneously the excitation functions with a sum of a Breit-Wigner and a second order polynomial function corresponding to the bound state signal and background, respectively. Moreover, the isospin relation between $n \pi{}^{0}$ and $p \pi{}^{-}$ pairs was taken into account. The corresponding data analysis is presented in detail in Ref.~\cite{Skurzok_NPA}.
The analysis resulted in the value of the upper limit in the range from 2.5 to 3.5~nb for the $dd\rightarrow$ ($^{4}\hspace{-0.03cm}\mbox{He}$-$\eta)_{bound}  \rightarrow$ $^{3}\hspace{-0.03cm}\mbox{He} n \pi^{0}$ process and from 5 to 7~nb for the $dd\rightarrow$ ($^{4}\hspace{-0.03cm}\mbox{He}$-$\eta)_{bound}  \rightarrow$ $^{3}\hspace{-0.03cm}\mbox{He} p \pi^{-}$ reaction. Systematic uncertainty, contributed mainly from the assumption of the Fermi momentum 
of the $N^{*}$ resonance inside $^{4}\hspace{-0.03cm}\mbox{He}$ 
~\cite{Kelkar_2016_new}, to be equal to that of a nucleon in $^4$He \cite{Nogga}, 
varies from 42\% to 46\% for both reactions.

These experimental results are revisited in the next section in the light of a new 
theoretical model~\cite{Ikeno_EPJ2017} which reproduces the 
$d d \rightarrow \, ^4$He$\eta$ cross section data and with the same $\eta$-$^4$He optical potential predicts the cross sections 
for $ d d$ fusion with the formation of an $\eta$-mesic $^4$He below the $\eta$ production threshold. The objective of the present analysis is twofold: to provide (i) stronger 
constraints on the optical potential parameters which are already capable of 
reproducing the $\eta$ production data and (ii) to improve the upper limits on the 
cross sections found in \cite{Skurzok_NPA} using a theoretical model (constrained by 
the above threshold data) for the possible bound state, rather than the simple 
Breit-Wigner form used in \cite{Skurzok_NPA}.

\section{Comparison between theory and data: results and discussion}~\label{Sec_3}
As mentioned in the previous section, 
we performed the analysis which allows to compare excitation functions measured for \mbox{$dd\rightarrow$ $^{3}\hspace{-0.03cm}\mbox{He} n \pi^{0}$} and \mbox{$dd\rightarrow$ $^{3}\hspace{-0.03cm}\mbox{He} p \pi^{-}$} processes~\cite{Skurzok_NPA} with the theoretical predictions presented in Ref.~\cite{Ikeno_EPJ2017}. For this purpose, theoretical conversion spectra were convoluted with the experimental resolutions of the excess energy $Q$. The COSY beam is characterized by a high momentum resolution of up to $\frac{\Delta p}{p}\approx 1\cdot10^{-4}$ resulting in the resolution of $\Delta Q$ of about 70~keV in the energy range of interest. This is about 70 times smaller than the binning of the spectra used by the WASA-at-COSY collaboration~\cite{Skurzok_NPA}. Hence, we bin the theoretical predictions in the same way as data, dividing the spectra into 20 intervals each of 5~MeV width. We assume also that the reconstruction efficiency in the WASA-at-COSY experiment is in a good approximation independent of the excess energy $Q$ as was proven in Ref. ~\cite{Skurzok_PhD}. An example of the theoretical spectrum after the binning procedure is presented in Fig.~\ref{theory_total_norm} as a black histogram.    

\begin{figure}[h!]
\centering
\includegraphics[width=9.0cm,height=6.5cm]{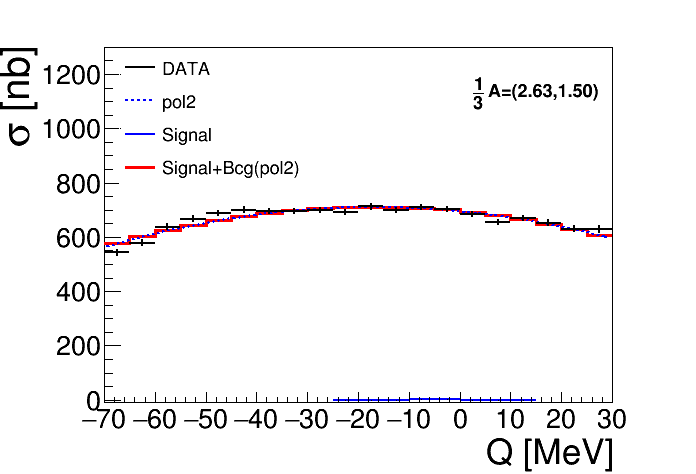}
\vspace{0.5cm}
\includegraphics[width=9.0cm,height=6.5cm]{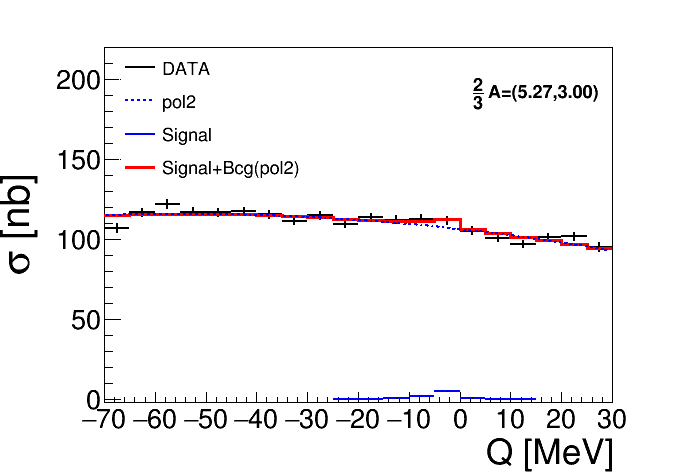}
\caption{Excitation function for \mbox{$dd\rightarrow$ $^{3}\hspace{-0.03cm}\mbox{He} n \pi^{0}$} (upper panel) and \mbox{$dd\rightarrow$ $^{3}\hspace{-0.03cm}\mbox{He} p \pi^{-}$} reaction (lower panel) determined as described in Ref.~\cite{Skurzok_NPA}. The red solid line represents a fit with
theoretical prediction for potential parameters ($V_{0},W_{0}$)=-(70,20)~MeV combined with a second order polynomial. The blue dotted line shows the second order polynomial describing the background while blue solid line shows the signal contribution. The experimental data~\cite{Skurzok_NPA} are indicated with black squares.~\label{exc_fit}}
\end{figure}

In the next step, the experimental excitation functions for \mbox{$dd\rightarrow$ $^{3}\hspace{-0.03cm}\mbox{He} n \pi^{0}$} and \mbox{$dd\rightarrow$ $^{3}\hspace{-0.03cm}\mbox{He} p \pi^{-}$} reactions were fitted simultaneously with a sum of binned theoretical function (signal) and a second order polynomial (background). The $n \pi^{0}$, $p \pi^{-}$ isospin relation was taken into account. The fitting functions can be presented as follows:

\begin{equation}
 \sigma_{n\pi^{0}}(Q)=\frac{1}{3}A\cdot Theory(Q) + B_{1}Q^{2}+C_{1}Q + D_{1}
\end{equation}
\begin{equation}
\sigma_{p\pi^{-}}(Q)=\frac{2}{3}A\cdot Theory(Q) + B_{2}Q^{2}+C_{2}Q + D_{2}
\end{equation}

\noindent for \mbox{$dd\rightarrow$ $^{3}\hspace{-0.03cm}\mbox{He} n \pi^{0}$} and \mbox{$dd\rightarrow$ $^{3}\hspace{-0.03cm}\mbox{He} p \pi^{-}$}, respectively. $Theory(Q)$ denotes the theoretical function after binning with the amplitude normalized to unity, while $B_{1,2}Q^{2}+C_{1,2}Q + D_{1,2}$ is a polynomial of the second order. The fit was performed for theoretical spectra obtained for different optical potential parameters ($V_{0},W_{0}$)~\cite{Ikeno_EPJ2017}. During the fit, the amplitude $A$ of the theoretical spectrum and polynomial coefficients were treated as free parameters. As an example, the excitation functions with the fit results for optical potential parameters ($V_{0},W_{0}$)=-(70,20)MeV are presented in Fig.~\ref{exc_fit}.

The performed fit delivers the amplitudes $A$ for \mbox{$dd\rightarrow$ $^{3}\hspace{-0.03cm}\mbox{He} N \pi$} consistent with zero within 2$\sigma$ for all sets of $V_{0},W_{0}$ parameters, which is given in Table~\ref{table1}. 
Therefore, the upper limit of the total cross section was determined, like in Ref.~\cite{Skurzok_NPA}, at the confidence level 90\% based on standard deviation of the amplitude $\sigma_{A}$ ($\sigma^{CL=90\%}_{upp}$=1,64$\cdot\sigma_{A}$). 
$\sigma^{CL=90\%}_{upp}$ values are presented for different parameters $V_{0},W_{0}$ in Table~\ref{table1}.\\
\begin{table}[h]
\begin{center}
\begin{tabular}{|c|c|c|c|}\hline
$V_{0}$  &$W_{0}$  & A (fit) [nb] &$\sigma^{CL=90\%}_{upp}$ [nb]\\
\hline 
-30 & -5  &-5.0$\pm$3.9  &6.5\\
-30 & -20  &-2.2$\pm$3.5 &5.8\\
-30 & -40  &0.2$\pm$3.8 &6.3\\

-50 & -5  &0.1$\pm$3.8 &6.3\\
-50 & -20  &3.3$\pm$4.1 &6.8\\
-50 & -40  &6.0$\pm$4.2 &6.9\\

-70 & -5  &6.4$\pm$4.5 &7.4\\
-70 & -20  &7.9$\pm$4.5 &7.4\\
-70 & -40  &7.5$\pm$3.7 &6.1\\

-100 & -5  &6.3$\pm$4.5 &7.4\\
-100 & -20  &6.9$\pm$3.9 &6.4\\
-100 & -40  &5.3$\pm$3.1 &5.2\\

\hline
\end{tabular}
\end{center}
\caption{Results obtained from the fit of theoretical spectra to experimental data. Table includes: optical potential parameters (first and second columns), amplitude obtained from the fit with its statistical uncertainty (third column) and upper limit of the total cross section for the $dd\rightarrow$ ($^{4}\hspace{-0.03cm}\mbox{He}$-$\eta)_{bound} \rightarrow$ $^{3}\hspace{-0.03cm}\mbox{He} N \pi$ process at CL=90\% (fourth column).~\label{table1}}
\end{table}

\begin{figure}[h!]
\centering
\includegraphics[width=8.5cm,height=5.5cm]{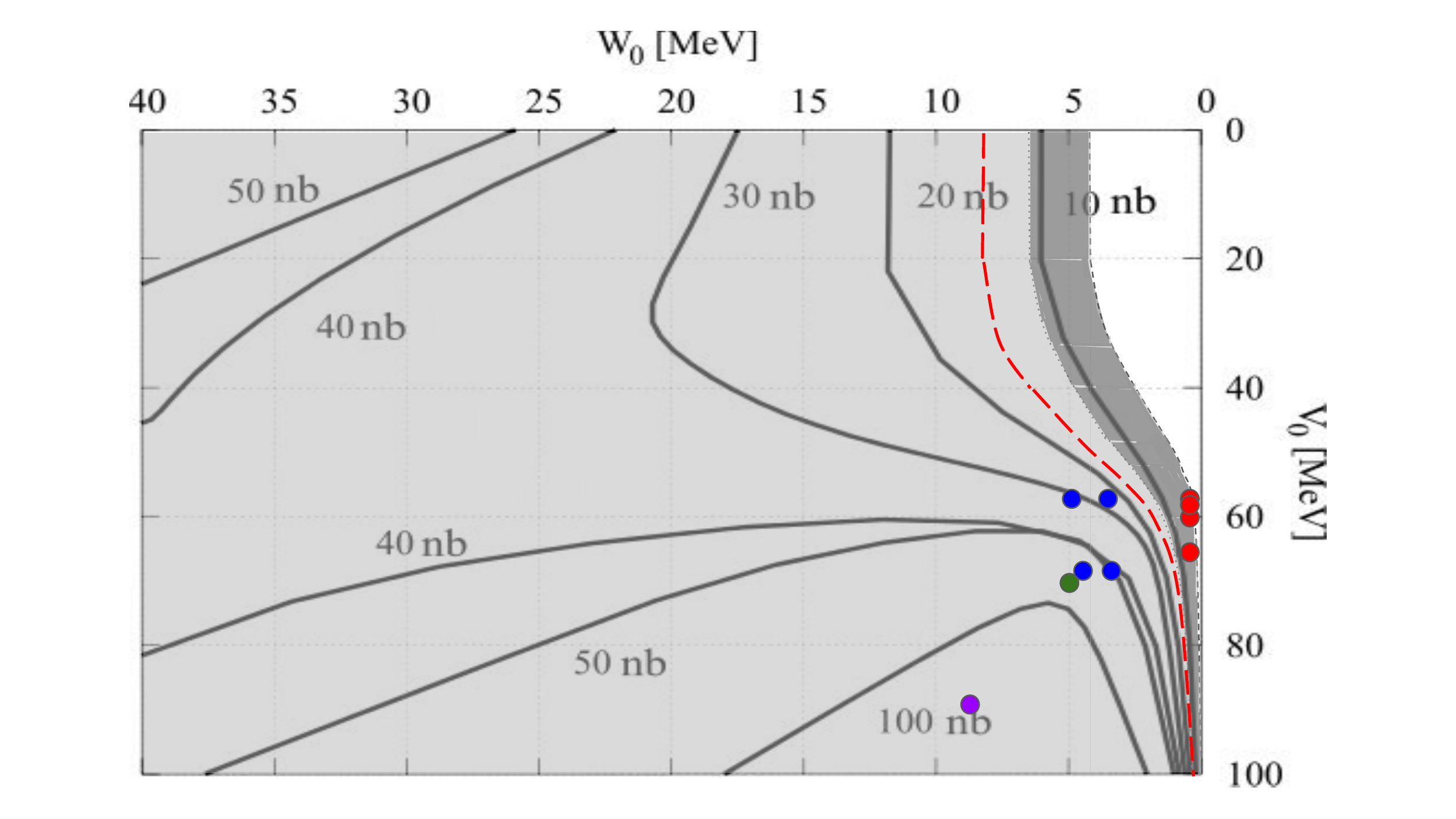}
\caption{Contour plot of the  theoretically determined conversion cross section in $V_{0} - W_{0}$ plane~\cite{Ikeno_EPJ2017}. Light shaded area shows the region excluded by our analysis, while the dark shaded area denotes systematic uncertainty of 
the $\sigma^{CL=90\%}_{upp}$. The red line extends the allowed region 
based on a new estimate of errors (see text for details). Dots correspond to the optical potential parameters corresponding to the predicted $\eta$-mesic $^4$He states.~\label{contour}}
\end{figure}

Obtained $\sigma^{CL=90\%}_{upp}$ is weakly sensitive to the $V_{0},W_{0}$ parameters, varying from 5.2 to 7.4 nb. Taking into account the systematic uncertainties of about 44\% estimated in Ref.~\cite{Skurzok_NPA}, the values increase, varying from about 7.5 to 10.7~nb. Therefore, in the contour plot shown in Fig.~\ref{contour}, we exclude the region where the cross section is above 10.7~nb (light shaded area). Dark shaded area shows the systematic error. The latter estimate is based on a calculation \cite{Kelkar_2016_new} for the $N^*$ 
momentum distribution for a given set of $\pi N N^*$ and $\eta N N^*$ coupling constants. 
If we take into account the calculations in \cite{Kelkar_2016_new} using 
all available values of the coupling constants, the 
allowed region in the $V_0$-$W_0$ plane can extend as far as the red line shown in 
Fig.~\ref{contour}. The coloured dots shown in the figure are the results of some optical 
model calculations which will be discussed in the next section.

\section{Optical model predictions of $\eta$-mesic $^4$He}

After constraining the region of the optical potential ($V_0$, $W_0$) parameter space 
allowed by the cross section data below the $\eta$ production threshold, let us now 
examine the possibility for the existence of $\eta$-mesic helium nuclei 
predicted within the optical model. 
To start with, we notice that all states predicted in Table 1 of 
\cite{Ikeno_EPJ2017} by solving the Klein Gordon equation with the 
optical potential of the present work, are excluded. 
On the other hand, since a wide range of 
$V_0$, $W_0$ values in \cite{Ikeno_EPJ2017} 
do reproduce the $d d \rightarrow \eta \,^4$He data, it seems worthwhile 
to investigate other optical model predictions in literature.

The authors in \cite{Gal2017} for example, 
compare their results using a few body formalism with existing optical model 
calculations by using the following form of the $\eta$-$^4$He potential with 
the complex $\eta$-nucleon scattering amplitude $F_{\eta N}$ chosen from two 
different models in literature \cite{GW,CS}:
\begin{equation}
V(r)=-{6 \pi \over 
\mu_{\eta N}}\,F_{\eta N}\,(r_0 \sqrt{\pi})^{-3}
\exp(-{r^2\over r_0^2}).
\label{eq:KMT}
\end{equation}
Replacing the parameter, $r_0= 1.267$ fm, as given in \cite{Gal2017} and
rewriting the above equation for the potential as,
$V(r) = 
[V_0 + i W_0] \, \exp(-{r^2}/{r_0^2})$, we identify the strengths $V_0$ and
$W_0$ and list them in Table~\ref{table2} for the different cases listed in Table 4 of Ref. \cite{Gal2017}.
\begin{table}[h]
\begin{center}
\begin{tabular}{|c|c|c|c|c|c|}\hline
$\eta N$ model  &$\delta\sqrt{s} [MeV] $ & $B_{\eta4He}$ [MeV] &
$\Gamma$ [MeV]&-$V_0$ [MeV] &-$W_0$ [MeV] \\
\hline
GW~\cite{GW} &0 &25.1 &40.8 &175.7&54.2 \\
 &-32.4 &1.03 &2.35 &89.7&8.6 \\
CS~\cite{CS} &0 &6.39 &21 &125.87&29.35 \\
 &-19.2 &- &- &69.15&5.046 \\
\hline
\end{tabular}
\end{center}
\caption{Strength of the optical potentials corresponding to the
$\eta$-$^4$He states given in \cite{Gal2017}.
$\delta \sqrt{s} = \sqrt{s} - \sqrt{s_{th}}$ with $\sqrt{s}$ being the
energy available in the center of mass of the $\eta N$ system.
$B_{\eta4He}$ and $\Gamma$ are the binding energies and widths of the
$\eta$-$^4$He states.
\label{table2}}
\end{table}
The $\eta N$ amplitude of \cite{GW} (GW), was 
obtained within a K-matrix description of the $\pi N$, $\pi \pi N$, $\eta N$ and 
$\gamma N$ coupled channels. The authors fitted the $\pi N \to \pi N$, 
$\pi N \to \eta N$, $\gamma N \to \pi N$ and $\gamma N \to \eta N$ data in the energy 
range of about 100 MeV on either side of the $\eta$ threshold. Ref. \cite{CS} 
presented the $\eta N$ amplitudes calculated within a chirally motivated separable 
potential model with the parameters of the model fitted to $\pi N \to \pi N$ and 
$\pi N \to  \eta N$ data. 
A comparison of the $V_0$ and $W_0$ values in Table~\ref{table2} with the allowed
region of the
$V_0 - W_0$ plane leads us to the conclusion that all the bound states
listed in Table~\ref{table2} are excluded by our analysis.

Having excluded the optical potential predictions of unstable bound states in literature, 
we turn to examine the special 
case of an unstable state centered at zero energy. 
The case of a zero energy bound state (or zero energy 
resonance), sometimes referred to as a transition state
\cite{barnea2} has been widely studied 
in literature \cite{deloff} in the context of different physical situations and 
has also been observed in ultracold atoms \cite{ultracold}. 
Let us recall some basic facts: a bound state corresponds to a pole in the 
S-matrix for $E < 0$. A resonance 
corresponds to a pole at positive energies. A state at $E = 0$ 
(which is usually referred to as a zero energy bound state in case the 
angular momentum $l >0$ and zero energy resonance otherwise) leads to a
scattering length, 
$a \to \infty$, i.e., the scattering length has a pole when $E =0$.
Ref. \cite{barnea2} has examined the occurrence of such states for a class 
of potentials of the form 
$V(r,r_0) = - {g \over r^s} \, f\biggl ( {r\over r_0} \biggr ) \, \, 
(g > 0, \,\,r_0 >0) $, 
which include the Gaussian, exponential and Hulthen among others. 
For the Gaussian optical potential of the present work, we identify 
$g$ with $V_0$, $s=0$ and $f = \exp({-r^2/r_0^2})$. 
Analytical as well as numerical solutions of the Schroedinger equation 
for these potentials are provided in Ref. \cite{barnea2}. It is shown 
that the existence of the transition state solution depends on a 
critical parameter given by 
$\beta = 2 \mu \, V_0\, r_0^2/ \hbar^2$, numerical values of which are 
listed in a table for several values of $l$. Taking their value of 
$\beta = 2.684$ in case of the Gaussian potential with $l = 0$, $\mu$ 
the reduced mass of $\eta$-$^4$He and with 
$r_0 = 1.267$~fm, we find $V_0 = -68.04$~MeV. 
Putting back this value in the expression, 
$V_0 = -[6 \pi/ \mu_{\eta N}]\,\Re e F_{\eta N}\,(r_0 \sqrt{\pi})^{-3}$, arising from 
(\ref{eq:KMT}), we determine $\Re e F_{\eta N} = 0.364$~fm. 
This value of $\Re e F_{\eta N}$ corresponds to the subthreshold energies 
of $\sqrt{s}$ = 1418.2 and 1467~MeV of the GW and CS $\eta$N amplitudes 
respectively (see Fig. 1 in Ref. \cite{Gal2017}). The imaginary parts 
of the amplitudes can be seen from the same figure in \cite{Gal2017} 
(at the corresponding energies) to 
be $\Im m F_{\eta N} = 0.0167$~fm and $\Im m F_{\eta N} = 0.0245$~fm for 
the GW and CS models respectively. The imaginary part of the optical 
potential can now be determined using, 
$W_0 = -[6 \pi/
\mu_{\eta N}]\,\Im m F_{\eta N}\,(r_0 \sqrt{\pi})^{-3}$.  

Thus, in case of the zero energy resonance, we find the optical 
potential parameters, ($V_0$, $W_0$) to be (-68.04, -3.12)~MeV and 
(-68.04, -4.55)~MeV for the GW and CS $\eta$-nucleon interactions respectively. 
Repeating the exercise for a different value of the Gaussian parameter, 
$r_0 = 1.373$~fm as in \cite{Ikeno_EPJ2017}, the potential parameters are 
found to be (-58.01, -3.2)~MeV and
(-58.01, -4.9)~MeV for the GW and CS $\eta$-nucleon interactions respectively. 

The above method of first considering the E=0 state of a real Gaussian 
potential to determine $V_0$ and then finding $W_0$ seems a posteriori 
justified considering the small values of $W_0$ (as compared to $V_0$) 
obtained. Indeed a similar procedure of first finding the binding energy  
by considering only the real part of the potential and later finding 
$\Gamma = - 2 < \Psi|\Im m V_{\eta A}|\Psi>$ using perturbation theory where 
$\Psi$ is the solution of the real Hamiltonian has been used in 
\cite{Gal2017} too.   

Finally, motivated by the above discussion, a renewed search for the $\eta$-$^4$He 
states within the model of \cite{Ikeno_EPJ2017} is performed. 
At the edge of the allowed region in Fig.~\ref{contour}, very narrow and 
weakly bound states of $\eta$-$^4$He, with binding energies and widths in the 
range of $\sim$ 2 - 230~keV and $\sim$ 8 - 64~keV respectively are found by 
solving the Klein Gordon equation as in \cite{Ikeno_EPJ2017}. These states 
correspond to the optical potential parameters $|V_0|$ in the range from 
58 to 65~MeV and $W_0$ = 0.5~MeV (red dots in Fig.~\ref{contour}). 
For values of $|V_0| <$ 58~MeV, no bound states are found.  
We should mention here, however, 
that some of the potential parameters are not consistent with the experimental data 
on the $\eta$ production cross section above threshold as reported in Ref. 
\cite{Ikeno_EPJ2017},  
especially for the cases with weak absorption. 
Hence we think that a systematic analysis including both the escape and 
the conversion cross sections covering the above- and below-threshold region
is necessary in order to investigate the weak absorptive potential region.

\section{Subthreshold considerations and uncertainties}
The $\eta$-nucleus optical potentials are in principle energy dependent and 
would depend strongly for example 
on the energy at which the elementary $\eta$N amplitude, $F_{\eta N}$, 
of Eq. (\ref{eq:KMT}) is evaluated. 
In the case of $\eta$-mesic nuclei, the $\eta$N interaction happens at 
subthreshold energies and $F_{\eta N}$ should be evaluated at an energy shifted 
by an amount $\delta$ below threshold. The importance of taking such a downward shift
into account has been discussed with different points of view in literature 
\cite{CieplyNPA2014,Galarxiv2017,HaiderIJMP,Hoshino}. The authors in 
\cite{CieplyNPA2014,Galarxiv2017} (and references therein) provide a detailed analysis 
of this topic and derive an expression for $\delta$ which depends on the nuclear 
binding energy per nucleon as well as the real part of the optical potential itself.
Refs. \cite{HaiderIJMP,Hoshino}, however, 
provide a simpler method with $\delta$ given by the average binding of the 
target nucleons.

\begin{figure}[h!]
\centering
\includegraphics[width=12.0cm,height=10cm]{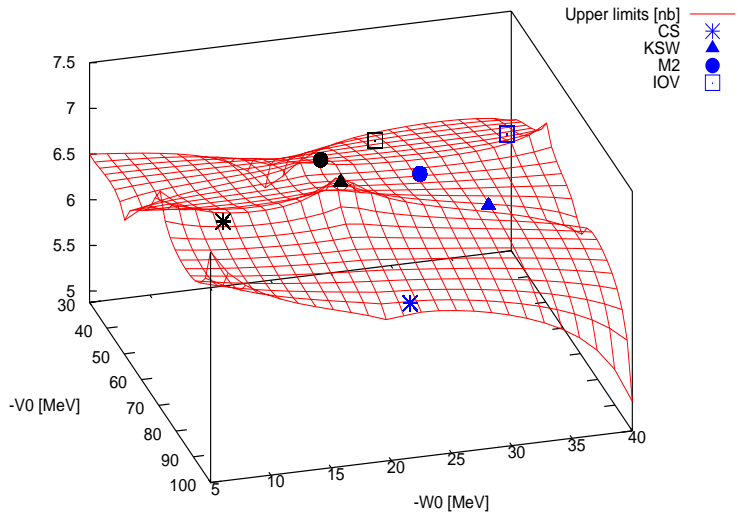}
\caption{Upper limits on the cross sections, $\sigma_{upp}$ in nb, 
as a function of the optical potential parameters $V_0$ and $W_0$. 
The red mesh represents the values determined in the present analysis 
(as in Table~\ref{table1}). The symbols are the values of $\sigma_{upp}$ corresponding 
to $V_0$, $W_0$ given in Table~\ref{table2} for the different $\eta$N models. The black 
symbols correspond to $\sigma_{upp}$ for $V_0$, $W_0$ evaluated using 
$F_{\eta N}$ at a subthreshold $\eta$N centre mass energy of $\sqrt{s}$ - 7 MeV and 
the blue symbols with $F_{\eta N}$ at threshold.
~\label{potsandsigma}}
\end{figure}

Since the experimental analysis of the present work relies on the input from the 
theoretical calculations in \cite{Ikeno_EPJ2017} where the above effects were not 
taken into account explicitly, we shall now try to estimate the uncertainties 
on $\sigma_{upp}$ (shown by the red mesh in Fig.~\ref{potsandsigma}) 
introduced by this omission. To obtain this estimate, we 
evaluate the optical potential parameters $V_0$, $W_0$ using Eq. (\ref{eq:KMT})
by comparing them with the form $V(r) = [V_0 + i W_0] \exp(-{r^2\over r_0^2})$. 
Thus, as observed in the previous section, 
$V_0 = -[6 \pi/ \mu_{\eta N}]\,\Re e F_{\eta N}\,(r_0 \sqrt{\pi})^{-3}$ and 
$W_0 = -[6 \pi/ \mu_{\eta N}]\,\Im m F_{\eta N}\,(r_0 \sqrt{\pi})^{-3}$. 
Evaluating $F_{\eta N}$ at threshold and at 7 MeV 
(binding energy per nucleon for $^4$He) and 30 MeV below threshold, 
we obtain the optical potential parameters given in Table~\ref{table2} for different 
models of $F_{\eta N}$ \cite{GW,CS,Mai2,KSW,IOV} in literature. 
The values of $\sigma_{upp}$ corresponding to the parameters $V_0$, $W_0$ in Table~\ref{table2} are read off from Fig.~\ref{potsandsigma} and listed in Table~\ref{table22}. Even if the optical 
potential parameters do change a lot depending on the choice of the energy at 
which $F_{\eta N}$ is evaluated, the upper limits on the cross sections do not seem 
to be very sensitive to this change. Depending on the model, the change in the 
upper limits can be between 0 - 6 \%. 
Given that the upper limits $\sigma_{upp}$ 
determined in the present analysis are not very sensitive to the 
parameters $V_0$, $W_0$ (see Table~\ref{table1} and Fig.~\ref{potsandsigma}), 
such a small uncertainty was expected. 

\begin{table}[h]
\begin{center}
\begin{tabular}{|c|c|c|c|c|c|c|c|c|c|}\hline
$F_{\eta N}$ &$\delta$=0 & & & $\delta$=-7& & &$\delta$=-30 & & \\
\hline
 &-$V_{0}$  &-$W_{0}$  & $\sigma_{upp}$& -$V_{0}$  &-$W_{0}$  & $\sigma_{upp}$ 
&-$V_{0}$  &-$W_{0}$  & $\sigma_{upp}$ \\
\hline
CS~\cite{CS}&97.7 &21.9&6.5 & 72.5&10 &6.88 &44.3 &2.7 &-  \\
M2~\cite{Mai2}&54.9 &28.8 &6.6 &45.2 &22 &6.59 &26.6&13 & - \\
KSW~\cite{KSW}&68.4 &32.6 &6.57 &56.8 &22 &6.67 &38.7 &13 &6.5 \\
IOV~\cite{IOV}&42.8 &37.8 &6.55 &36.35 &27.8 &6.46 &20.16 &16.5 &-  \\
GW~\cite{GW}&139 &43.6 &- &104 &23.7 &- &71.7 &8.1 &6.95  \\
\hline
\end{tabular}
\end{center}
\caption{Optical potential parameters $V_0$ and $W_0$ (in MeV) 
evaluated using (\ref{eq:KMT}) with
the $\eta$-N amplitude $F_{\eta N}$ 
evaluated at $\delta =0$ (threshold), $\delta$ = -7 MeV 
and $\delta$ = -30 MeV, with $\delta = \sqrt{s} -\sqrt{s_{th}}$. 
The upper limits on the cross sections listed in this table are read from 
the mesh (representing the $\sigma_{upp}$ (in nb) 
determined in the present analysis) in Fig.~{\ref{potsandsigma}} at the values of 
$V_0$ and $W_0$ in this table. \label{table22}}
\end{table}

\section{Summary and Conclusions}

We performed an analysis in order to constrain the $\eta$-$^4$He 
optical potential parameters by comparing a recently developed theoretical 
model for $\eta$-$^{4}\hspace{-0.03cm}\mbox{He}$ bound state production 
in \mbox{$dd\rightarrow$ $^{3}\hspace{-0.03cm}\mbox{He} N \pi$} reactions~\cite{Ikeno_EPJ2017} with the experimental data collected by WASA-at-COSY~\cite{Skurzok_NPA}. 
Convoluting the theoretical cross section with experimental resolutions,  
we estimated the upper limits of the total cross sections for the formation of the $\eta$-mesic Helium nuclei in \mbox{$dd\rightarrow$ $^{3}\hspace{-0.03cm}\mbox{He} N \pi$} processes at a 90\% confidence level. 
Comparison of the determined upper limits for the creation of $\eta$-mesic nuclei via the \mbox{$dd\rightarrow$ $^{3}\hspace{-0.03cm}\mbox{He} N \pi$} process with the cross sections obtained in Ref.~\cite{Ikeno_EPJ2017} excludes a wide range of $\eta$-$^{4}\hspace{-0.03cm}\mbox{He}$ optical potential parameters. With the values of $|V_0|$ and $|W_0|$ being 
restricted to be less than 60 MeV and 7 MeV respectively, most predictions of 
$\eta$-mesic helium states seem to be excluded within the present analysis. 
Extremely narrow and loosely bound states within the model of \cite{Ikeno_EPJ2017} 
seem however to appear in the 
allowed region of the optical potential parameters. 

In spite of some shortcomings such as the absence of the explicit inclusion of the 
strong energy dependence of the $\eta$N interaction \cite{Galarxiv2017} 
and the fact that, in principle, the $\eta$-helium nuclei should be treated within a few 
body formalism \cite{Fix2017,Gal2017,RakitPRC,KelkarPRL}, it is worth noting that 
in the decades long search 
for $\eta$-mesic nuclei, the present work is indeed a first 
attempt to combine the experimental data below $\eta$ production threshold with 
predictions from a theoretical model which can reproduce the existing data above 
threshold too. There exist approaches such as the coupled channels 
generalization of the optical potential~\cite{WycechKrzemien} which can bring 
out interesting aspects related to the existence of the $\eta$-mesic helium.
Hence, it is hoped that the optical
model analysis of the present work should provide guidance in narrowing down
the search for $\eta$-mesic $^{4}\hspace{-0.03cm}\mbox{He}$.


\section{Acknowledgements}
We acknowledge the support from the Polish National Science Center through grant No. 2016/23/B/ST2/00784 and the Faculty of Science, Universidad de los Andes, 
Colombia, through project number P17.160322.007/01-FISI02. This work was partly supported by JSPS KAKENHI Grant Numbers JP16K05355 (S.H.) and  JP17K05443 (H.N.) in Japan.


\section*{References}

\bibliography{mybibfile}

\end{document}